\documentstyle[preprint,aps,epsfig]{revtex}

\def\be{\begin{equation}}
\def\ee{\end{equation}}

\begin{document}
\draft
\title{Neutrino Star as Galactic Cluster Dark Halo}
\author{M.~H.~Chan and M.~C.~Chu}
\address{Department of Physics, The Chinese University of Hong Kong, \\
         Shatin, New Territories, Hong Kong, China. }

\maketitle

\begin{abstract}
Recent strong lensing data and rotation curves of dwarf
galaxies indicate that many galactic clusters may have a
soft core instead of a central cusp in their density distribution. This
result challenges the standard CDM (Cold Dark Matter) theory prediction
based on N-body simulations.
We find that the observed density profile is consistent with that of
a spherical gas of degenerate fermions such as neutrinos
in hydrostatic equilibrium.
Also, we compare two different models of the dark matter
halo and their predictions for the hot gas profile.

\end{abstract}
\vskip 5mm

Recent strong lensing data and rotation curves of dwarf galaxies suggest
that matter distribution takes the form of a soft core at the
centers of some galaxy clusters \cite{Tyson,Firmani}. However,
N-body simulations based on the CDM theory predict that the
density profile of the dark matter halo should be singular at the
center which is called the NFW profile \cite{Navarro,Navarro2}.
One possibility to resolve the discrepancy is that the dark matter
may not be collisionless as assumed in previous simulations.
They may interact with each other and form a profile different
from the NFW profile. In the following, we show that indeed the
observed density profile is consistent with that of a degenerate gas
of massive neutrinos in hydrostatic equilibrium.
We will first review a method of inferring the dark matter distribution
from the density profile of the hot gas in a galaxy cluster
\cite{Jones,Thomas}.
Then we will demonstrate that one of the dark matter candidates,
massive neutrinos, which can be either collisionless or self-interacting, can
account for the observed profiles.

One of the most common methods to probe the dark matter distribution is to
make use of the gas density profile. We assume that the hot gas is in
hydrostatic equilibrium, i.e.,
\be
\frac{dn_g(r)}{dr}=- \frac{GM(r)m_gn_g(r)}{kTr^2} ,
\label{hotgas1}
\ee
where $n_g(r)$ is the number density of hot gas, $M(r)$ is the total mass
enclosed within a radius $r$, $k$ is the Boltzmann constant, and
$m_g$ is the mass of the gas particles. Here we assume that the hot gas is
isothermal throughout the cluster, with a constant temperature $T$, which
is valid  as the heat conduction by electrons is quite efficient.
By knowing the number density profile of the hot gas, we can probe the
distribution of dark matter. 
There is a general form of the hot gas density profile called the $\beta$
model \cite{Jones},
\be
n_g(r)= \frac{n_0}{[1+{\tilde r}^2]^{3 \beta /2}} ,
\label{hotgas2}
\ee
where ${\tilde r} \equiv r/r_c$, $r_c$ being the core radius, 
$\beta$ is a parameter,
and $n_0$ is the central number density.
Different clusters have different values of $n_0$, $r_c$ and $ \beta$.
Now, we can solve Eq.~(\ref{hotgas1}) and Eq.~(\ref{hotgas2}) and get:
\be
M(r)= \frac{3kT \beta r^3}{Gm_g(r^2+{r_c}^2)} .
\label{totalmass}
\ee
For $r$ much greater than $r_c$, $M(r) \propto r$, whereas for $r$ much smaller
than $r_c$, $M(r) \propto r^3$, which means that the density
is constant,
\be
 \rho \rightarrow \rho _c \equiv\frac{ 9kT\beta}{{4 \pi Gm_g r_c}^2}\ , 
\ \  r<<r_c.
\label{rhodm1}
\ee
In general, from Eq.~(\ref{totalmass}), we can obtain the mass density of
the dark matter, assuming that the total mass is dominated by dark matter
together with the hot gas,
\be
 \rho_{\rm DM}= {\rho _c \over 3}\left[ \frac{{\tilde r}^2+3} 
{({\tilde r}^2+1)^2} \right]-m_gn_g(r) \ , \label{rhodm2}
\ee
which is quite different from the NFW profile,
\be
 \rho_{\rm DM}= \frac{ \rho_0}{(r/r_s)(1+r/r_s)^2} \ ,
\label{rhodm3}
\ee
where $\rho_0$ and $r_s$ are parameters.
Possible sources of the discrepancy are
problems in the NFW profile, or the $ \beta$ model, or the
assumption of isothermal distribution in the cluster.
For example, it has been proposed that including the cooling
flow of hot gas in the above model will result in a dark matter
distribution that is closed to the NFW prediction \cite{Adami}.

Nevertheless, the hot-gas method is model dependent and the uncertainties
are still large. To understand more about the dark matter,
one powerful and model independent alternative is to use
strong lensing to probe the distribution of dark matter. This method
still has large errors and is limited in
data collection.  One of the clusters studied recently is CL0024+1654
\cite{Tyson}. It was concluded that CL0024+1654 has a soft core
in the center which is different from the NFW
profile \cite{Shapiro}. Furthermore, some observations of
dwarf galaxies mass distribution also indicate that the inner halo density
is nearly constant \cite{Firmani}. 

It has been proposed that cold dark matter may
have weak self-interaction instead of being collisionless \cite{Spergel}.
Following this suggestion, we assume that the dark matter 
can establish hydrostatic equilibrium,
\be
 \frac{dP_{\rm DM}}{dr}=- \frac{GM(r) \rho_{\rm DM}(r)}{r^2} ,
\label{hystat1}
\ee
where $P_{\rm DM}$ is the dark matter pressure.
Then, we can combine Eq.~(\ref{hotgas1}) and Eq.~(\ref{hystat1}) to get
\be
 \frac{dP_{\rm DM}( \rho_{\rm DM})}{ \rho_{\rm DM}}= \frac{kT}{m_g}
\frac{dn_{g}}{n_{g}} .
\label{hystat2}
\ee
From Eq.~(\ref{hystat2}), we can obtain the equation of state (EOS) of 
dark matter $P_{\rm DM}( \rho_{\rm DM})$ from the hot gas density profile.
If we assume a polytropic relation between the pressure and density of the
dark matter,
\be
P_{\rm DM}=K {\rho_{\rm DM}}^{ \gamma} \ ,
\label{darkeos}
\ee
where $K$ and $\gamma$ are parameters in the EOS, Eq.~(\ref{hystat2}) 
becomes
\be
 \ln{ \left( \frac{ \rho_g}{ \rho_{gc}} \right)}= \frac{m_g \gamma K}{kT(
\gamma -1)} \left[{ \rho_{\rm DM}}^{ \gamma-1}-{ \rho_{\rm DMc}}^{
\gamma-1} \right],  \label{rhog}
\ee
with $ \gamma>1$, and $ \rho_{gc}$ and $ \rho_{\rm DMc}$ are central
density of hot gas and dark matter respectively. 
The hot gas density profile directly tracks the dark matter profile and is
sensitive to the steepness of the dark matter EOS. 

In the following, we consider massive neutrinos as a specific model of dark 
matter and solve the cluster core problem. 
At low density, neutrinos are collisionless as they have very small
interaction cross-section. However,
at high density they become degenerate, and they can therefore be in
hydrostatic equilibrium. It has been proposed that such neutrino ``stars''
may be found in the centers of galaxies \cite{Viollier,Apparao}.
Here we propose that they may exist at
the centers of galaxy clusters.
Since the interaction cross-section of neutrinos is small, we can treat them
as a zero temperature gas. 
The degenerate pressure for non-relativistic neutrinos is given by
\be
P_{ \nu}= \frac{h^2}{5m_{ \nu}} \left ( \frac{3}{8 \pi} \right )^{2/3}{n_{
\nu}}^{5/3} \ ,
\label{nueos}
\ee
where $n_\nu$ and $m_\nu$ are the number density and mass of the neutrinos.
We can then solve for the density profile of a neutrino star assuming
hydrostatic equilibrium (shown in Fig.~1), which is expectedly
similar to that of a white dwarf. Here we set the neutrino mass to be 3 eV
and the central number
density to be $10^{8}$~cm$^{-3}$, which fits the observational data for
clusters. The radius of the neutrino star is several kpc which is 
comparable to the cluster scale. The total mass is about $10^{14}$ solar 
masses, which is also close to the total mass of a cluster.
A neutrino star has a core in the center, the
radius of which depends on the central mass density only. The higher the
central mass density is, the smaller is the core radius and vice versa.
The inner density profile is similar to the data obtained from 
gravitational lensing. This suggests that at least some clusters may be 
degenerate fermions dominated.
From Fig.~1, we can approximate the neutrino star density profile as
following: for $r<<r_0$, $M(r)=Cr^3$, where $C$ is a constant, and
for $r>>r_0$, $M(r)=M_0$, where $M_0$ and $r_0$ are the total mass
and core radius of the neutrino star.
Using the above approximation, we can analyse the density profile of the
hot gas inside a cluster core if its mass is dominated by the neutrino
star. If $M(r)=Cr^3$, we have
\be
n_g=n_0e^{-ar^2} ,
\label{ng1}
\ee
where $a= GCm_g/2kT$.  For $r<<a^{-1/2}$, the hot gas profile is smooth 
and flat.  Therefore, the core radius is
\be
r_c=a^{-1/2}= \left( \frac{2kT}{GCm_g}\right)^{1/2} .
\label{rc}
\ee
Clearly, $r_c$ is not necessarily equal to $r_0$, but it is affected by the
central mass density of the neutrino star and temperature of the hot gas.
In our model, for $T=10^8$~K and central mass density 
$\rho_c=10^{-25}$~gcm$^{-3}$, $r_c$ is about 150~kpc.

We can also calculate the density profile of the hot gas numerically. In
Fig.~2, we show the hot gas profile using the above parameters of a
neutrino star, and we find that it is well fitted by using a $\beta$ model
with $ \beta=1.1$ and $r_c$=150~kpc. This indeed agrees with the observational
value of $\beta$ between 0.9-1.1 \cite{Markevitch1,Markevitch2}.

We now compare the results of using three different
models, the NFW dark matter profile, neutrino star model and hot gas
self-bounded profile.
Here, we continue to assume that the temperature of the hot gas is
constant throughout the cluster. Fig.~3 shows the hot gas profiles from
these three models. We can see that the three hot gas profiles have
similar central mass density and
temperature. However, the slopes of the hot gas
profiles for $r>r_c$ are different, being smallest for self-bounded model
and largest for the neutrino star model.
In fact, the slope of the hot gas profile depends not only on the dark
matter profile but also on the temperature profile of the hot gas and
velocity
dispersion of the dark matter particles. Therefore, the dark matter
profile obtained by using the hot gas profile alone is model
dependent. For $r>r_0$, the total density profile is
dominated by the hot gas and the total density tail $ \rho_T$ is
approximately proportional to $r^{-3}$ which agrees with the observational
result \cite{Kneib}.  To understand the dark matter distribution better,
gravitational lensing is a potentially more powerful method. 
It is model independent, and we can know the mass density
profile directly without using any hot gas model if the cluster is 
dominated by the dark matter.  For example, the cluster 
CL0024+1654 observed by Tyson 
$et$ $al.$ does not have a cD galaxy, which means that the mass of 
dark matter is dominant at the center of the cluster \cite{Firmani}. 
Gravitational lensing data suggests that the NFW profile may not be correct.

We have shown in this article that observational data on cluster hot gas is
consistent with the existence of neutrino stars, which can account for at 
least parts of the dark matter.  But how are these neutrino stars formed?
It is commonly believed that
there is a cosmological neutrino background originating from 
the Big Bang \cite{Peebles}, 
corresponding to a temperature of about 1~K now, or about 0.1~meV 
in energy scale. Therefore it is conceivable that cosmological neutrinos with 
rest mass even in the sub-eV range are non-relativistic.  Recent 
experimental observations of neutrino oscillations point to a range of 
the mass difference between muon and tau neutrinos to be 
$5 \times 10^{-4}< \Delta m^2<6 \times 
10^{-3}$ eV$^2$ ~\cite{Fukuda}.  Therefore at least one species of the 
cosmologocal background neutrinos should be non-relativistic, and
the interplay between gravitational attraction and degenerate pressure
makes it possible to create neutrino stars.
We have carried out hydrodynamics simulations of their formation process,
and preliminary results indicate that hydrostatic neutrino
stars can indeed form with a wide range of parameters \cite{manho}. 

One of the interesting properties of the neutrino star is that
it has a flat core, which agrees with the gravitational lensing data 
as well as the rotation curves of dwarf galaxies, but contradicts
with the NFW profile. Recently, Sand $et$ $al.$ has provided an upper 
limit of the slope of the central density 
profile $\alpha<0.57$ which is different from NFW's 
($\alpha=1$) \cite{Sand}. In 
our model, $\alpha$ is approximately zero. In order to solve 
the "core problem" of galactic clusters, more data will clearly be needed.

The work described in this paper was substantially supported by a grant
from the Research Grants Council of the Hong Kong Special Administrative
Region, China (Project No. 400803).

\newpage
\begin{figure}[h]
\psfig{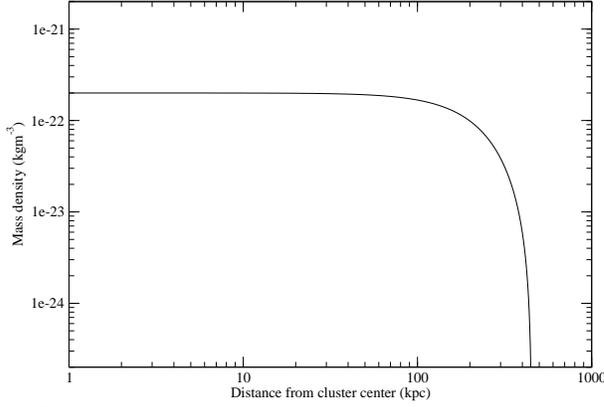}
\caption{The mass density profile of a neutrino star
with central mass density $ \rho_{\nu}=10^{-25}$~gcm$^{-3}$ with a neutrino
mass of 3~eV.}
\end{figure}

\begin{figure}[h]
\psfig{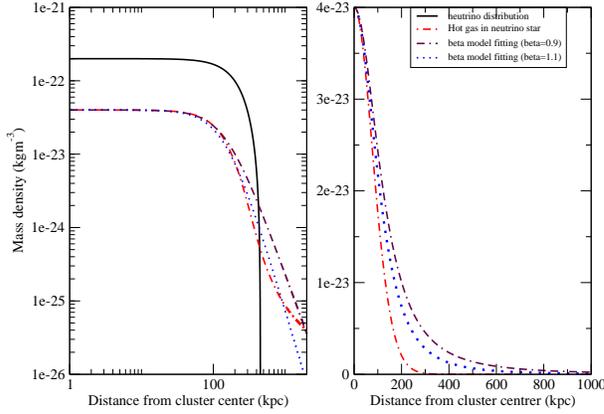}
\caption{Numerical result of the hot gas profile and beta model
fitting with T=$10^8$~K, $r_c$=150~kpc and $n_g$=20000 at the center.
Right panel: hot gas profile together with beta model fitting.  Left panel:
the central core of the hot gas in log scale; dark matter profile is also 
shown (solid line).}
\end{figure}

\begin{figure}[h]]
\psfig{file=cluster3b.eps,angle=0,width=8cm}
\caption{Hot gas profiles in four
different models: NFW model from CDM simulation, with $r_s$=300~kpc (dot-dashed
line), neutrino star model (solid line), beta model (dashed line),
and self-bounded model (dotted line). For the latter three models,
same parameters as in Fig.~2 have been used.}
\end{figure}

\begin{figure}[h]]
\psfig{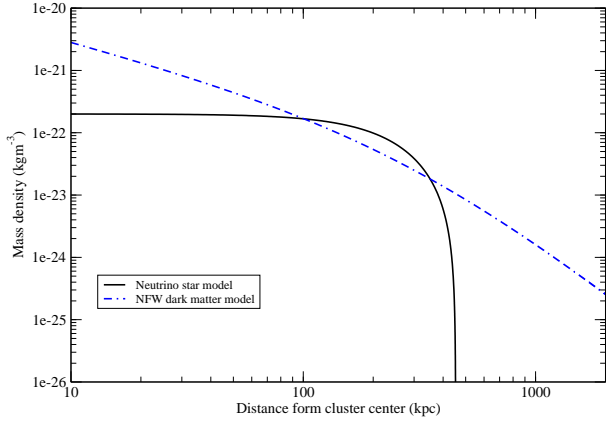}
\caption{Dark matter profiles in neutrino star model (solid line) vs.~NFW
model from CDM simulation, with $r_s$=300~kpc (dot-dashed line).
Same parameters as in Fig.~2 have been used.}
\end{figure}

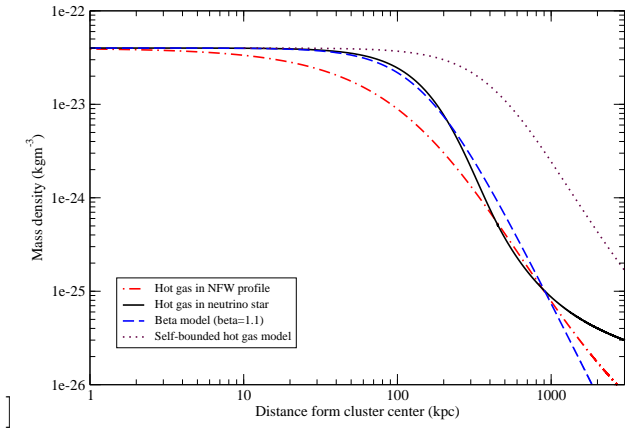
\end{document}